\newcommand{\ba}{\begin{array}}
\newcommand{\ea}{\end{array}}
\newcommand{\bean}{\begin{eqnarray*}}
\newcommand{\eean}{\end{eqnarray*}}
\newcommand{\rm}[1]{\mathrm{#1}}
\newcommand{\tr}{\mathrm{tr}}
\newcommand{\Tr}{\mathrm{Tr}}

\newcommand{\bra}[1]{\left\langle {#1} \right|}
\newcommand{\ket}[1]{\left|  #1 \right\rangle}

\newcommand{\braket}[2]{\left\langle #1 | #2 \right\rangle \;}

\newcommand{\bmx}[1]{\left(\begin{array}{*{#1}{c}}}
\newcommand{\emx}{\end{array}\right)}
\newcommand{\bmxw}[1]{\renewcommand{\arraystretch}{2}\left(\begin{array}{*{#1}{c}}}
\newcommand{\bmxww}[1]{\renewcommand{\arraystretch}{2.5}\left(\begin{array}{*{#1}{c}}}
\newcommand{\bdet}[1]{\renewcommand{\arraystretch}{1.2}
	\left|\begin{array}{*{#1}{c}}}
\newcommand{\edet}{\end{array}\right|\renewcommand{\arraystretch}{1}}
\newcommand{\beq}{\begin{equation}}
\newcommand{\eeq}{\end{equation}}
\newcommand{\bea}{\begin{eqnarray}}
\newcommand{\eea}{\end{eqnarray}}

\newcommand{\ditem}[1]{\item[$\diamond$]}

\newcommand{\bit}{\begin{itemize}}
\newcommand{\eit}{\end{itemize}}

\newcommand{\eab}{\begin{eqnarray}}
\newcommand{\eae}{\end{eqnarray}}

\documentclass[aps,prb,twocolumn,superscriptaddress]{revtex4-1}
\usepackage{graphics}
\usepackage{epsfig}
\usepackage{amssymb}
\usepackage{color} 
\usepackage{subfigure}
\begin{document}
%\begin{fmffile}{fgraphs}

% Use the \preprint command to place your local institutional report
% number in the upper righthand corner of the title page in preprint mode.
% Multiple \preprint commands are allowed.
% Use the 'preprintnumbers' class option to override journal defaults
% to display numbers if necessary
%\preprint{}

%Title of paper
\title{Topological Insulator and the $\theta$-vacuum in a system without boundaries}

% repeat the \author .. \affiliation  etc. as needed
% \email, \thanks, \homepage, \altaffiliation all apply to the current
% author. Explanatory text should go in the []'s, actual e-mail
% address or url should go in the {}'s for \email and \homepage.
% Please use the appropriate macro foreach each type of information

% \affiliation command applies to all authors since the last
% \affiliation command. The \affiliation command should follow the
% other information
% \affiliation can be followed by \email, \homepage, \thanks as well.
\author{Kuang-Ting Chen}
%\homepage[]{Your web page}
%\thanks{}
%\altaffiliation{}
\affiliation{Department of Physics, Massachusetts Institute of Technology, Cambridge, MA 02139}
\author{Patrick A. Lee}
%\homepage[]{Your web page}
%\thanks{}
%\altaffiliation{}
\affiliation{Department of Physics, Massachusetts Institute of Technology, Cambridge, MA 02139}

%Collaboration name if desired (requires use of superscriptaddress
%option in \documentclass). \noaffiliation is required (may also be
%used with the \author command).
%\collaboration can be followed by \email, \homepage, \thanks as well.
%\collaboration{}
%\noaffiliation

\date{\today}

\begin{abstract}
In this paper we address two questions concerning the effective action of a topological insulator in one and three dimensional space without boundaries, such as a torus. The first is whether a uniform $\theta$-term with $\theta=\pi$ is generated for a strong topological insulator. The second is whether such a term has observable consequences in the bulk. The answers to both questions are positive, but the observability in three dimension vanishes for infinite system size.

\end{abstract}

% insert suggested PACS numbers in braces on next line
%\pacs{
% insert suggested keywords - APS authors don't need to do this
%\keywords{}

%\maketitle must follow title, authors, abstract, \pacs, and \keywords
\maketitle

% body of paper here - Use proper section commands
% References should be done using the \cite, \ref, and \label commands
\section{Introduction}
The topological insulators are characterized by insulating band structures with nontrivial topology, which cannot be smoothly deformed back to an atomic insulator, sometimes under certain discrete symmetry.\cite{clkane,liangfu,moorebalents} The most well-known example is the integer quantum Hall effect (IQHE) in two dimensions (2D). Here despite the fact that the bulk of the system is gapped, the system possesses $n$ gapless chiral edge states where $n$ is the number of occupied Landau levels. In addition, if we perturb the system using local electric fields, there will be local transverse current in the system. This effect is best captured by the Chern-Simons term $\int \rm d^2x\rm d t\epsilon^{\mu\nu\lambda}A_\mu \partial_\nu A_\lambda$ in the effective theory. 

In three dimensions (3D), under time-reversal (TR) symmetry there is a $Z_2$ classification for band insulators,\cite{liangfu,moorebalents} which distinguishes the usual insulator from the topological insulator. In contrast to the IQHE states, the two classes of insulators are often distinguished only on the edge: in the bulk both insulators do not have a current response to the applied field. However, there is an odd number of spin-filtered gapless edge states, i.e., helical Dirac cones, on the surface of the topological insulator. Another physical effect of the topological insulator can be observed when the TR symmetry is locally broken on the edge: the Dirac cones would be gapped and result in an $\frac12$-integer quantum Hall effect on the edge.\cite{QHZ} This effect also leads to the bulk electromagneto-polarizability.\cite{AJD}

The effect mentioned above can also be understood using the effective topological field theory.\cite{QHZ} This theory postulates that the band topological insulator can be described by a bulk "$\theta$-term", 
\beq
\mathcal{L}_\theta\equiv\frac{\theta e^2}{32\pi^2}\epsilon^{\mu\nu\lambda\omega}F_{\mu\nu} F_{\lambda\omega}, \label{ltheta}
\eeq
with $\theta=\pi$. The integrand in eq. (\ref{ltheta}) is a total derivative so the equation of motion remains unaltered in the bulk. The edge quantum Hall effect can be derived with a smooth change of $\theta$ from $\pi$ to $0$ in space. Finally, the gapless spin-filtered edge states, are understood as a limit when one preserves the TR symmetry on the boundary.

There is, however, some confusion as to whether the bulk $\theta$-term, given by eq. (\ref{ltheta}) is indeed required to describe the topological insulator. Specifically, the "$\nabla\theta$-term",
\beq
\mathcal{L}_{\nabla\theta}\equiv-\frac{ e^2}{16\pi^2}\partial_u\theta\epsilon^{\mu\nu\lambda\omega}A_\nu F_{\lambda\omega},\label{lgtheta}
\eeq
which differs from eq. (\ref{ltheta}) by a total derivative, is what actually has been derived in the previous theories and can produce all the effects mentioned above as well. There is an important conceptual distinction between the two theories: with the former theory, it looks as if the bulk of the topological insulator has some special property, which manifests itself either on the edges or when the Hamiltonian is varied; with the latter theory, the bulk of the topological insulator is not intrinsically different from an ordinary insulator, and all the possible physical consequences only occur at the edge between it and a normal insulator, or with a variation of the Hamiltonian.

In this paper, first we show that there are physical consequences which distinguish between the two theories, and then we show that it is the former theory, $\mathcal{L}_\theta$, which describes the topological insulator. The paper is organized as follows: in the following section, we study the effective theory on a closed manifold and show that the two theories given by eq. (\ref{ltheta}) and eq. (\ref{lgtheta}) behave differently. In section III, we start from the fermionic band structure and derive the effective theory $\mathcal L_\theta$, without the ambiguity of a total derivative.

\section{The physical consequence of the $\theta$ term}

Without a boundary, $\mathcal{L}_{\nabla\theta}$ will have no effect if $\theta$ is uniform. Therefore, the distinction between the two aformentioned theories will be evident if there is any physical consequence of a $\theta$-term with uniform $\theta$. In the following we shall discuss the effect of an uniform $\theta$-term in one and three dimensions, with various topologies. 

%\begin{enumerate}

%\item 
\textit{$\theta$-term in one spatial dimension (1D):}

Here we shall follow the approach of the $\theta$-vacuum, where the $\theta$ term gives a prescription to form gauge-invariant states. Please see appendix A for a derivation directly using the path integral.

For completeness we first show that the $\theta$-vacuum description is equivalent to a path integral with $\mathcal L_\theta$. The $\theta$-term in 1D is defined as 
\beq
\mathcal{L}_{\theta,1D}=\frac{e\theta}{2\pi}\epsilon^{\mu\nu} \partial_\mu A_\nu.
\eeq

 Let us first take the $A_0=0$ gauge. Define $\tilde A_1(q)=\int A_1\exp(-iqx)\mathrm{d}x$ as the Fourier transform of $A_1$. On a circle of circumference $L$, configurations satisfying $\int A_1\rm dx\equiv\tilde{A}_1(0)=0$ can be gauge transformed into configurations satisfying $\tilde{A}_1(0)=2\pi n/e$, with $n$ an integer (the winding number). Therefore, when we consider a state that is an eigenstate of the quantized operator $\tilde{A}_1(0)$, say, with eigenvalue $0$, we should consider instead a linear combination of all states, each with eigenvalue $2\pi n/e$. The linear combination has to be gauge invariant, and the remaining arbitrary choice would be the phase between states with consecutive $n$. We call this relative phase $\theta$ and call the vacuum of this phase the $\theta$-vacuum:
\beq
\ket{\theta,\text{phys}}=\sum_n\exp(-i\theta n)\ket{n,\text{phys}}.
\eeq 

Notice that if we write down the path integral from some state with winding $n$ to some other state with winding $m$ by turning on $A_1(t)$, the winding number can be written as an integral:
\beq
m-n=\frac{e}{2\pi}\int^m_n\rm dx\rm dt\frac{\partial A_1}{\partial t};
\eeq
here the limits of the integral denote the winding number of the initial and final configuration. The vacuum-vacuum amplitude can thus be expressed as
\bea
&&\sum_{m,n}\bra{m,0}\exp(iHt)\ket{n,0}\exp(i\theta(m-n))\nonumber\\
&=&\sum_{m,n}\int_{n,0}^{m,0}[DA_1]\exp(iS+i\frac{e\theta}{2\pi}\int\rm dx\rm dt\frac{\partial A_1}{\partial t});
\eea
here $S$ in the exponent is just the ordinary action corresponding to $H$ and the scripts of the integral specifies the initial and final boundary conditions.
The $\theta$-vacuum description is thus equivalent to adding $\mathcal L_\theta$ to the Lagrangian.

Now we proceed to derive the physical consequence of the term. Consider a Maxwell Lagrangian with vacuum angle $\theta$ at finite temperature $1/\beta$. Taking $A_0=0$, the Maxwell Hamiltonian is
\beq
H=\sum_q\frac1{2L} \left|\frac{\partial\tilde{A}_1(q)}{\partial t}\right|^2=\sum_q\frac1{2L}|\tilde{E^1}(q)|^2\equiv\sum_qH_q
\eeq
Since $\tilde{A}_1(q\neq 0)$ decouples from $\tilde{A^1}(0)$ we can calculate them independently. $\theta$ only couples to the $q=0$ sector as all operators at finite $q$ have the same eigenvalue for states which differ by arbitrary winding. Let us focus on the partition function of the $q=0$ sector:
\begin{widetext}
\beq
Z_{q=0}=\Tr_\theta(e^{-\beta H_0})
=\frac{e}{2\pi}\int_0^{2\pi} \rm d\phi\int^\infty_{-\infty} \frac{\rm d\ell}{2\pi}\sum_m\sum_n\braket{\phi+2\pi m}{\ell}\bra{\ell}e^{-\frac{\beta Le^2}{2}\ell^2}\ket\ell\braket{\ell}{\phi+2\pi n}e^{i(m-n)\theta};
\eeq
\end{widetext}
the subcript $\theta$ denotes that we only trace over the sector whose vacuum is the $\theta$-vacuum. $\phi=e\tilde A(q=0,\tau=0)$ is the initial configuration of the gauge field. Note that we have inserted $1=\int^\infty_{-\infty} \frac{\rm d\ell}{2\pi}\ket{\ell}\bra{\ell}$, where $\ell$ is the eigenvalue of $(\tilde{E^1}(q=0)/eL)$ and $\ket\ell$ the eigenstate. The canonical conjugate pairs $(x,p)$ can be determined from the Lagrangian with $p=\frac{\partial \mathcal{L}}{\partial \dot x}$; if we choose $(e\tilde{A^1}(q=0))$ as $x$ it conjugates to $(\tilde{E^1}(q=0)/eL)$. Therefore we have 
\beq
\braket{\phi+2\pi m}{\ell}=\exp(i(\phi+2\pi m)\ell).
\eeq
There is translational symmetry in $m$ and $n$ and the sum over $m+n$ just gives an overall normalization constant. If we replace $(m-n)$ by $n$, we have
\beq
Z_{q=0}=\int^\infty_{-\infty} \frac{\rm d\ell}{2\pi}\sum_ne^{in(\theta+2\pi\ell)}e^{-\frac{\beta Le^2}{2}\ell^2}.
\label{z1d}
\eeq
If we sum over $n$ first, we have
\beq
\sum_ne^{in(\theta+2\pi\ell)}\sim \sum_m\delta(\frac{\theta}{2\pi}+\ell+m).
\eeq
Physically, this means the effect of the uniform $\theta$ term is to cause the average electric field to be quantized in integer units of charges, but shifted by $e\theta/2\pi$. This is a well-known result with open boundary conditions\cite{coleman}, where one can imagine fractional charges at the end produce the electric field. With periodic boundary conditions it is less intuitive.

%With $\theta=\pi$, the electric field can only take half-integer values. Also note that, perhaps rather unexpected, even with $\theta=0$ the electric field on a circle is still quantized. This is a consequence of the gauge invariance, as the large gauge transformation effectively compactifies the $\tilde A(0)$ field. In contrast, in a vacuum with fixed winding number, the electric field going through need not be quantized. Neverthess, this vacuum is not gauge invariant subject to those large gauge transformations.

If $\theta=\pi$, this would imply that the vaccum has two degenerate configurations characterized by $\frac1L\int\rm dxE=\pm\frac12 e$. The matrix elements between the two states become exponentially small as $L\rightarrow\infty$, so we should think of this as a sponteneous symmetry breaking situation where the parity (P) and charge-conjugation (CC) symmetry are sponteneously broken by the electric field. The electric field would choose one direction and stay for a time period proportional to $e^L$.

In conclusion, there is indeed a real measurable difference between $\mathcal{L}_\theta$ and $\mathcal{L}_{\nabla\theta}$, where with $\theta=\pi$ in the former theory there will be huge ground state electric field at $q=0$ and the CC symmetry is sponteneously broken, whereas in the latter theory there will be no effect and the symmetry is preserved.

%\section{Topological Insulator in Three Dimensions} 
%Similarly, the symmetry-breaking edge response of topological insulators in %three dimensions can be captured by a term $\frac{\theta(x)}{2\pi} %\vec{E}\cdot\vec{B}.$ However, in the setting without edges, we can still %distinguish between this effective action and the "$\nabla\theta$" theory, the %integrated-by-part version. Here again we investigate if there is any %bservable effect in the former theory, with an uniform $\theta.$ Here we %consider three settings: (i) abelian gauge field on a 3-torus; (ii) abelian %gauge field on a 3-sphere and magnetic monopoles; (iii) nonabelian gauge field, %on a 3-sphere.

%\item 

Now we turn our attention to three dimensions. We consider two settings without boundaries: the first is a 3-torus, and the second is the 3-sphere. We restrain ourselves to consider only U(1) gauge fields.

\textit{Abelian gauge field on a 3-torus:} 

Since we need a periodic lattice to produce the topological insulator, it is natural first to consider the world as a 3-torus. Again taking the gauge choice $A_0=0$, the $\theta$-term in three spatial dimensions can be written as a difference of the Chern-Simons term on the initial and the final states in the imaginary time direction:
\bea
\int_S{\rm d}^4 x\mathcal{L}_\theta&=&\int_S \frac{{\rm d}^4 x}{8\pi^2} \epsilon^{\mu\nu\lambda\gamma}\partial_\mu A_\nu \partial_\lambda A_\gamma\nonumber\\
&=&\int_{\partial S}\frac{{\rm d}^3 x}{8\pi^2}A_i \partial_j A_k\epsilon^{ijk},
\eea
where $i, j, k$ now run through only the spatial directions. One superficial difference to the situation in 1D is that it seems all finite-$q$ components contribute. However, as we require the initial and final states to differ from each other only by a gauge transformation, $\vec A_{\rm final}=\vec A_{\rm initial}+\nabla\phi/e$, we can see the integral on the three-dimensional boundary becomes a total derivative,
\beq 
\int{\rm d}^3 x\partial_i (\phi\partial_j A_k\epsilon^{ijk}/e)=\int{\rm d}^3x\partial_i(\phi B_i).\label{3dp}
\eeq
Let us assume $\phi$ only has a winding in the $z$ direction, i.e., $\phi(x,y,L_z)-\phi(x,y,0)=2\pi n$, then eq. (\ref{3dp}) becomes
$2\pi n\Phi_B$ where $\Phi_B$ is the total flux threading the torus in the $z$ direction. Assuming $\Phi_B=m\Phi_0$ with $\Phi_0=hc/e=2\pi/e$, we find
\beq
\int_S \rm d^4 x \epsilon^{\mu\nu\lambda\gamma}\partial_\mu A_\nu \partial_\lambda A_\gamma= \frac{8\pi^2}{e^2}nm,
\eeq
Thus, with $m$ units of the fundamental flux quantum in the $z$ direction, the "$\theta$-vacuum" consists of linear superposition of states with configurations satisfying $\int A_z\rm dz=2\pi n/e$, where $n$ is an integer. Since the Hamiltonian is still quadratic, we can calculate the relevant part of the partition function similar to the calculation in 1D. The analog of eq. (\ref{z1d}) is
\begin{widetext}
\bea
Z_{q=0}&\sim&\sum_{m,n}\int\frac{\rm d\ell}{2\pi}\; e^{imn\theta}e^{i2\pi n\ell}\exp\left(-\frac{\beta V}{2}((\frac{e\ell}{L_xL_y})^2+(\frac{2\pi m}{eL_xL_y})^2)\right)\nonumber\\
&\sim&\sum_{m,n'}\int\frac{\rm d\ell}{2\pi}\delta(\frac{m\theta}{2\pi}+\ell+n')\exp\left(-\frac{\beta V}{2}((\frac{e\ell}{L_xL_y})^2+(\frac{2\pi m}{eL_xL_y})^2)\right).
\eea
\end{widetext}

$V=L_xL_yL_z$ is the world volume and we choose our conjugate variables to be $(e\int A_z\rm d^3x /L_xL_y)$ and $(\int E_z \rm d^3 x/eL_z)$, with the eigenvalue of the latter labeled by $\ell$. We find that with a fix flux $\Phi_B=m\Phi_0$ in the $z$ direction, the electric flux in the same direction is quantized: 
\beq
E_zL_xL_y=e(n-m\theta/2\pi)=ne-\theta\Phi_B/\Phi_0^2, 
\eeq
with $n$ an integer.

Let us first take the strict $T=0$ limit. Here the thermal fluctuation of the magnetic flux is suppressed and we find that, the $\theta$-term only has nontrivial effect if there is a finite flux threading through. For $\theta=\pi$, when we have an odd magnetic flux, the electric flux in the same direction would be quantized in half units of $e$. The electric field goes to zero if the world volume goes to infinity, however. 

At finite $T$, the thermal fluctuation of the magnetic field can generate some finite fluxes, and we would have some effect even with $B=0$ in average. For simplicity let us again set $\theta=\pi$ and consider $L_x=L_y=L_z=L$. If $T\gg 1/L$, The correlation function of the electric field would contain an extra term comparing to the usual Maxwell theory:
\beq
\langle E(x)E(y)\rangle\sim \langle E(x)E(y)\rangle|_{\theta=0}+\frac{e^2}{8L^4}.
\eeq

One can understand this constant correlation by imagining that half of the states in the ensemble have an odd number of magnetic fluxes. The state with an odd number of fluxes would have a ground state electric field squared to $(e/2L^2)^2$, and the average is just a half of that. This extra part of the correlation function is long ranged, and can easily be distinguished from the Maxell part. However, the magnitude again vanishes in the large $L$ limit. Since it is not possible to have a 3D torus without embedding it in 4D space, these effects are of academic interests only.

Before we end this subsection, we should note that from this calculation, it is clear that any local magnetic field will not produce any effect. Therefore, one would not see an electric field inside a solenoid, nor any charge at the end of it.

%Now we can ask the same question whether the 3D topological insulator is described by the $\theta$-term with $\theta=\pi$, under this boundary condition. Similar to the case in 1D, this amounts to the question whether the path integral from one winding to another has a phase proportional to both the threading magnetic flux and the difference of the winding number. It is easier to think about a large number of threading fluxes, where the electron then sits in some Landau levels. Note that the Landau levels only disperse in the direction along the magnetic field and one can choose a gauge such that the wave function are localized in the orthogonal directions, and have a 1D character. Therefore, from the claim we have for the 1D topological insulator, we can also claim that the $\theta$-term does not describe the 3D topological insulator. 

%Nevertheless, we note that unlike in 1D, the difference between the correlation function of the electric field, with or without the $\theta$-term, vanishes in the thermaldynamic limit. 

%\item 
\textit{Abelian gauge field on a 3-sphere and magnetic monopoles}: 

Since we cannot have global nonzero magnetic flux in any direction in a 3-sphere, there will be no effect of the $\theta$-term. This is in contrast to the case with a magnetic monopole, where it is predicted that there will be charge $e(n-\theta/2\pi)$ attached to it in a $\theta$-vacuum. This effect can be understood as follows: magnetic monopole is a singularity in terms of the abelian gauge field. Suppose we have a pair of monopole-antimonopole far away in a 3-sphere so that we have one fundamental flux going from one to the other. The geometry is now a 3-sphere with two punctures. From the calculation of the previous section we can see the electric flux threading from one hole to the other must be quantized, $\Phi_E=e(n-\theta/2\pi)$, and we would attribute this as the charge of the magnetic monopole.

Franz et. al. showed that there is Witten effect inside the topological insulator.\cite{franz} We emphasize here that this does not prove that a bulk $\theta$-term exists, as the Witten effect can also come from the "$\nabla\theta$" theory, provided that we characterize it by $\theta=0$ inside the monopole. Given that a monopole can only live in a unit cell, and the band structure is absent in the unit cell, it is not unnatural to set $\theta=0$ inside a monopole. 

As a side note, if we consider nonabelian gague fields, the $\theta$-term in general does have effect in a 3-sphere. This effect, however, is usually associated with the physics of instantons and is quite different from what we have discussed.

%\item 
%\textit{Nonabelian gauge field on a 3-sphere:}
 
%With nonabelian gauge fields, we no longer need a finite magnetic flux in the initial state to observe the $\theta$-vacuum. Nevertheless, the vacua with different winding numbers are now connected only via instantons and the process would be suppressed by $e^{-S_I}$, where $S_I$ is the instanton action. Furthermore, if we assume a Higgs mechanism to break the nonabelian gauge symmetry down to the $U(1)$ we observe at low energy, the instanton energy would always be of order $v^2$ where $v$ is the symmetry breaking scale. We conclude that physical effects in this sector is entirely different from that of the topological insulator.

%\end{enumerate}

\section{the effective $\theta$-term in the presence of band electrons}
In this section, we investigate how the presence of fermions can alter the vacuum $\theta$-angle. We review the topological band invariant which characterizes the topological insulators in one and three spatial dimensions and show that they are related to the shift of $\theta$. For the clearness of the formula, sometimes we take the units $e=1$; i.e., we absorb $e$ into the definition of the external electromagnetic gauge fields.

The topological insulator in 1D is characterized by the polarization\cite{KV} 
\beq
P=\int\frac{\mathrm{d}k}{2\pi}\sum_{\text{occ}}-i\bra{u_i}\frac{\partial}{\partial k}\ket{u_i}=\int\frac{\mathrm{d}k}{2\pi}\tr\left(\mathcal{A}_x\right);
\eeq
with
\beq
\mathcal{A}_{\mu,nn'}\equiv \bra{u_{nk}}-i\frac{\partial}{\partial k^\mu}\ket{u_{n'k}}
\eeq 
which is the so-called Berry's phase gauge field in momentum space. $\ket{u_{nk}}$ is the periodic part of the Bloch wave function. 

\begin{figure}[htb]
	\centering
	\subfigure[normal insulator]{\includegraphics[width=4cm]{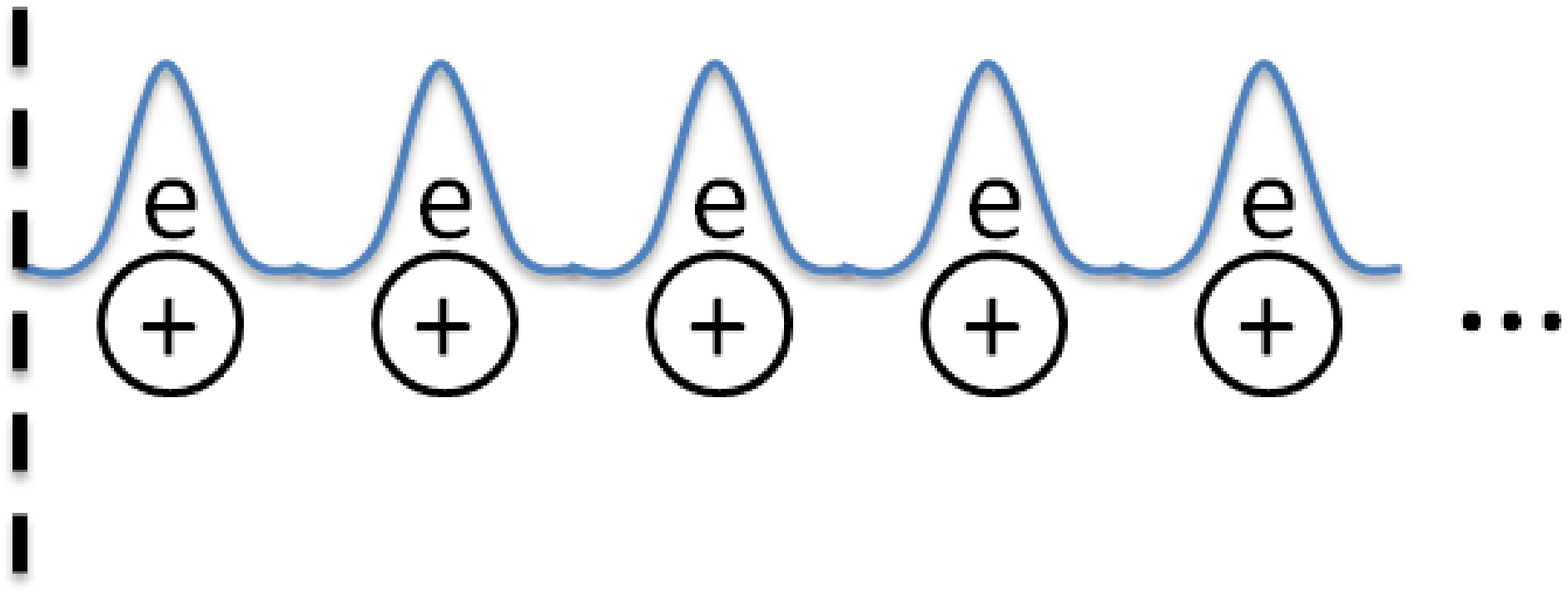}}
	\subfigure[topological insulator]{\includegraphics[width=4cm]{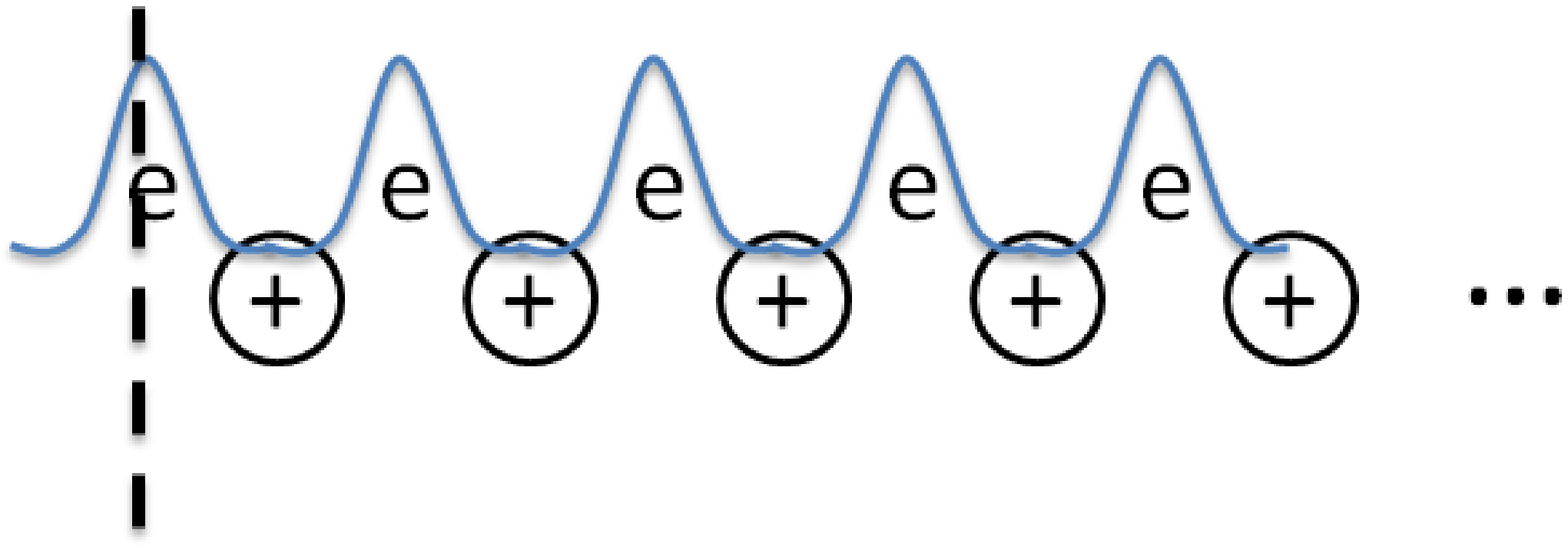}}
	\caption{Topological insulator in 1D. The dashed line shows the edge. The envolope shown is the Wannier wave function of the electrons.}
	\label{cartoon}
\end{figure}

$P$ is forced to either take the value $\frac12$ or $0$ when there is charge conjugation (CC) symmetry. If the CC symmetry is preserved everywhere, then on the boundary there will be an odd number of zero modes. If the CC symmetry is locally broken in some way on the boundary, then there will be a $n+1/2$ charge, where $n$ depends on the detail of the local symmetry breaking. A cartoon showing the effect is depicted in Fig. (\ref{cartoon}).

Naively one might think this already shows that $\theta$ is shifted by $2\pi P$: after all, the $\theta$-term in 1D is nothing but an energy term proportional to the electric field, in which the energy of dipoles, $-\int {\rm dx}P\cdot E$, fits perfectly. However, we should note that normally this dipole energy arises from separating the charges to the boundary, away from each other. On a circle with uniform polarization, therefore, one would not anticipate such energy term is present.

\begin{figure}[htb]
	\centering
	\subfigure[the procedure]{\includegraphics[width=4cm]{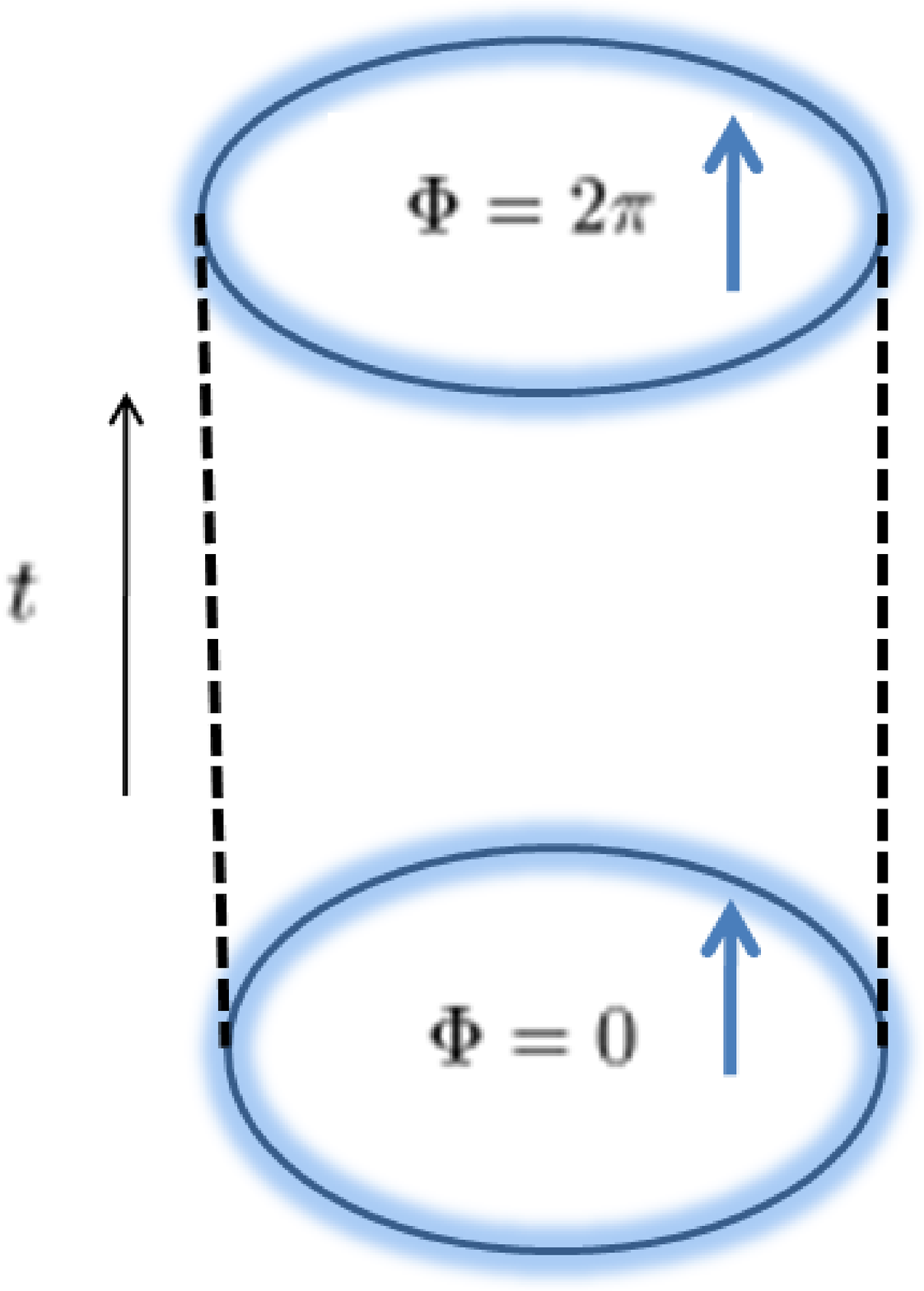}}
	\subfigure[spectral flow]{\includegraphics[width=4.5cm]{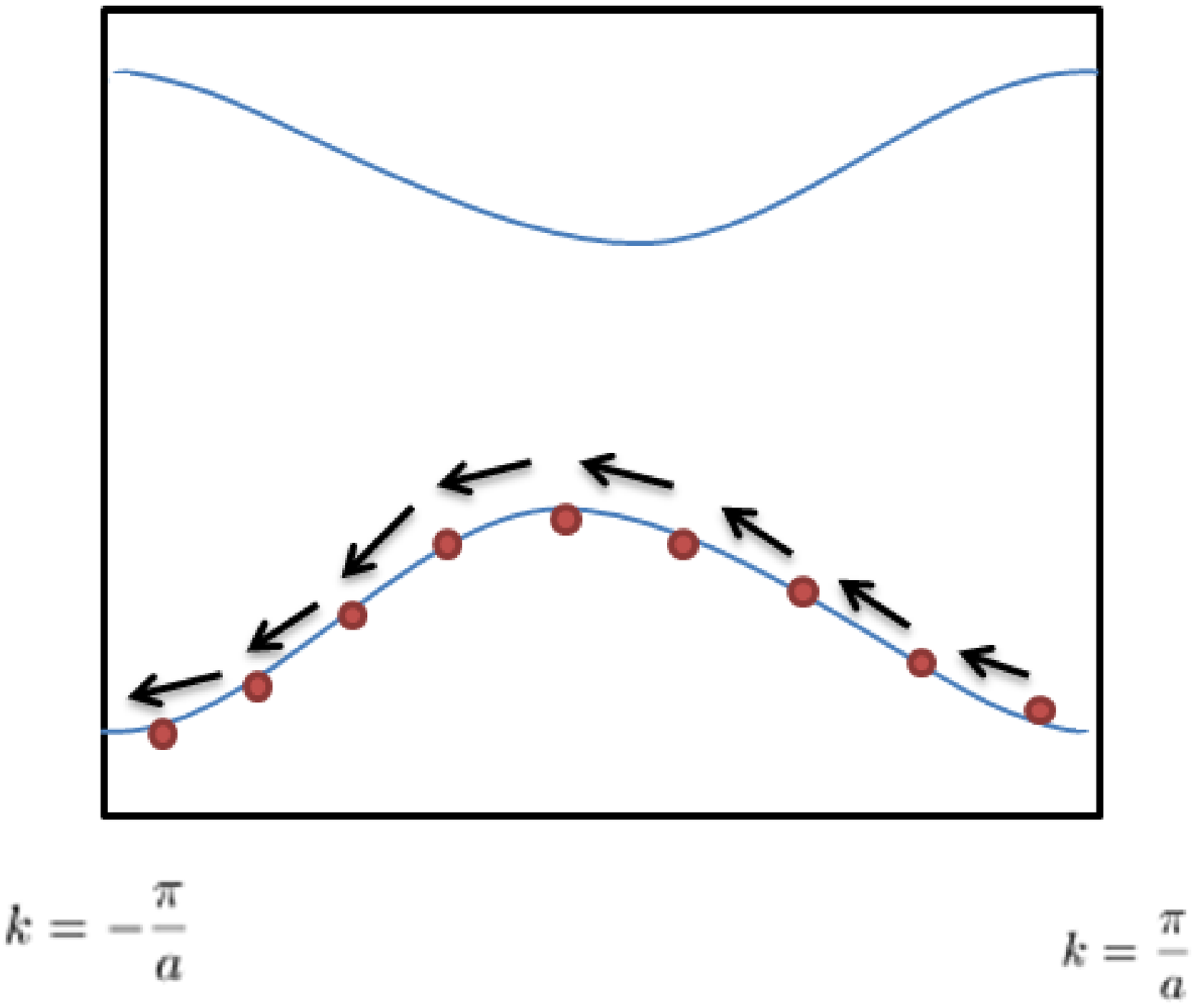}}
	\caption{(a) A flux is slowly threaded through. $\Phi=0$ and $\Phi=2\pi$ are the same physical state related by a gauge transform. We calculate the Berry's phase of the process. (b) During the process, at every allowed momentum by the periodic boundary conditions, the energy and the periodic part of the wave function moves slowly to the values of the state to the left, according to eq. (\ref{uflow}). When a full flux is threaded, each one of them takes the eigenvalue and the eigenvecotor of the one at its left. Note that the momentum quantum number $k$, however, does not change. When we sum over the Berry's phase contribution from all the single particle states, it becomes an integral over the entire Brillouin zone.}
	\label{1dprocess}
\end{figure}

If we look back at how $\theta$ change the physical property of the system, it comes in by adding a phase to the amplitude between vacuua with different winding numbers. Specifically, $\theta$ is precisely the additional phase of the amplitude between vacuua with consecutive winding numbers. In the presence of  gapped fermions, this phase can come from integrating out the fermions, in other words, the dynamical phase the fermionic system obtains under a time-dependent back ground field. This phase has two contributions, one is just the time-dependent energy of the fermions and the other is the Berry's phase of the process. The phaseshift from the energy depends on the time duration and is not just a function of the initial and final state; therefore it does not alter $\theta$. Therefore, similar to the consideration in Ref.\cite{RS}, we are led to consider the accumulated geometric phase of the band electrons, when the external field is slowly turned on.See Fig. \ref{1dprocess}(a) for a cartoon of the procedure.  

First we shall consider how the single particle wave function change as we increase $A_1$ uniformly. We have
\beq
\psi_{nk}(x)=u_{nk}(x)e^{ikx}
\eeq
which is the wave function in position basis, and $u_{nk}(x)$ is periodic and satisfies
\beq
\left(\left(\nabla-(k+eA_1)\right)^2+V(x)\right)u_{nk}(x)=E_{nk}u_{nk}(x).\label{uchange}
\eeq

As we increase $A_1$ uniformly to $A_1+\eta$, the momentum $k$ cannot change as it is fixed by the finite size $L$ and the periodic boundary condition. On the other hand, following eq. (\ref{uchange}), $u_{nk}(x)$ changes as
\beq
u_{nk}(A_1+\eta)=u_{n(k-e\eta)}(A_1)\label{uflow},
\eeq
which is just a corresponding shift of $k$ by $-e\eta$. if $e\eta=2\pi/L$, the system returns to its original state, but in a different gauge (i.e., with winding number different by one.) Notice that  while $u_{nk}(x)$ goes to the next avaiable value on the left, the $k$ in the exponential stays the same. The electronic wave function is therefore different from its starting state. Nevertheless, as discussed further below, if we include the gauge field, the final state differs from the initial state by a large gauge transformation, and the Berry's phase accumulated in the process is exactly what we want to calculate.

As a side note, the situation is similar if we put electrons on a lattice which couples to the gauge field via Periels substitution. The single particle eigenfunction can be written as $\psi_{nk}=\sum_i u_{nk,m}\exp(ikx_i)\ket{m,x_i}$, with $u_{nk}$ now a vector in the orbital space. With an increase of $A_1$, only $u_{nk}$ changes. 

Now we are ready to calculate the accumulated Berry's phase of the band electrons under the process, where the winding of the gauge field is increased by one:
\bea
\phi_\mathrm{Berry}&=&i\int_{2\pi n/e}^{2\pi(n+1)/e}\rm d\tilde A_1(0)\bra{\Psi_e}\frac{\partial}{\partial \tilde A_1(0)}\ket{\Psi_e}\nonumber\\
&=&i\int_{2\pi n/e}^{2\pi(n+1)/e}\rm d\tilde A_1(0)\sum_{k_i,\alpha\in\rm{occ}}\bra{\psi_{k_i\alpha}}\frac{\partial}{\partial \tilde A_1(0)}\ket{\psi_{k_i\alpha}}\nonumber\\
&=&i\sum_{k_i,\alpha\in\rm{occ}}\int_{k_i}^{k_i+2\pi/L}{\rm d}k\bra{u_{k\alpha}}\frac{\partial}{\partial k}\ket{u_{k\alpha}}\nonumber\\
&=&i\int_{\rm BZ}{\rm d}k\sum_{\alpha\in{\rm occ}}\bra{u_{\alpha k}}\frac{\partial}{\partial k}\ket{u_{\alpha k}}\nonumber\\&=&-2\pi P.
\eea
In the second equality, we wrote the derivative acting on the Slater determinent as a sum of derivatives acting on single particle wave functions. In the third equality we then plug in the dependence of the wave functions, and change variables to $k$. Whenever $\tilde A_1(0)$ increases by $2\pi/e$, each $u_{nk}$ reaches the next allowed eigenstate to the left by the periodic boundary condition (without actually changing the momentum eigenvalue.) As we sum over all the integral of eigenstates at different allowed $k$'s, the whole Brillouin zone (BZ) is covered exactly once and we reach the fourth equality. 

While we calculate the Berry's phase for process where the winding number of the initial and final states differs by one, evidently the phase is proportional to the difference of the winding number in general. Therefore, it leads to a shift of $\theta$ by $\phi_{\rm Berry}$. Specifically, if the vacuum has an vacuum angle $\theta=0$, we see that the topological insulator in 1D will be described by $\theta=\pi$. 

A few remarks are in order. Firstly, this calculation is good for a finite-size system, where both the lattice spacing and the length of space are finite. Despite that the eigenstates in such a system would be discrete points in the BZ, the whole BZ is covered by the integral and there is no finite-size effect. Secondly, one might notice that the fermionic wave functions of the initial and final state are different, and it seems that our process does not form a close loop as usually required by a physical (gauge-invariant) Berry's phase. This does not invalidate our calculation, however, since the initial state and the final state are related by a large gauge transform. Once the convention of the phase of the initial state is determined, the phase convention of the final state is also determined via gauge-transforming the initial state to have the desired winding number. The residual gauge degree's of freedom without altering the external gauge field (which is just an arbitrary phase of the final state) can be absorbed into the vacuum angle $\theta$ in the absence of the fermions.

Now we turn to the 3D strong topological insulator. The 3D strong topological insulator, defined under TR symmetry, is characterized by the band structure invariant\cite{QHZ}
\beq
\frac{1}{4\pi}\int\mathrm{ d}^3k\epsilon^{abc}\Tr(\mathcal{A}_a\partial_b\mathcal{A}_c-i\frac23\mathcal{A}_a
\mathcal{A}_b\mathcal{A}_c)=\pi.
\eeq

Similar to the case in 1D, a $\theta$-term is generated if we can find a Berry's phase of the electrons which is proportional to the difference of the winding number of the initial and final gauge configuration. More specifically, the procedure is as follows: we apply a constant finite magnetic field, say, in the $z$ direction on the 3-torus. We then slowly change the gauge field in the same direction uniformally until the final state is connected to the initial state by a large gauge transform. Then we apply the magnetic field in some other direction and repeat the calculation. We can also consider procedures such as applying magnetic field in $z$ direction and changing the gauge $y$ direction; the phase of this process leads to a term $\propto E_yB_z$ in the effective Lagrangian. In general, we therefore expect the full effective theory to take the form 
\beq
\mathcal L_{\rm eff}=\sum_{ij}\alpha_{ij}E_iB_j.
\eeq
The $\theta$-term is the isotropic part of the effective Lagraigian:
\beq
\theta=\frac{4\pi^2}{3e^2}\sum_i\alpha_{ii}.
\eeq  

Two things are different from our calculation in 1D: firstly, the state at a given $k_z$ is already a many-body wave function. Secondly, we have to calculate everything in a finite (but maybe small) magnetic field. In principle one can go over all the Laudau levels at a given $k_z$ and sum up their Berry's phases, but in practice this is not easy to do as the wave functions are not perturbative in $B$. It turns out that the density matrix perturbation theory introduced in Ref.\cite{AAJD} is suitable for this calculation with an extra trick as will be described below.

Before we dig into the calculation, let us clarify that our calculation, despite taking advantage of the same formalism, is distinct from Ref.\cite{AAJD}. There they first calculate the current flowing through the bulk as they vary the Hamiltonian with time under a small magnetic field, then they relate the time integral of the current to the polarization. While the uniform $\theta$-term in 3D with boundaries would give rise to a magneto-polarization effect, the converse cannot be said. As mentioned in the introduction, both $\mathcal{L}_\theta$ and $\mathcal{L}_{\nabla\theta}$ can produce this effect, so a derivation of the effect does not distinguish between the two theories. To our best knowledge the following calculation is the first demonstrating that it is indeed $\mathcal{L}_\theta$ which describes the topological insulator.

Suppose we apply a small magnetic field along the $z$-direction, $\vec B=B\hat z$. Following the calculation in 1D, we calculate the Berry's phase of the process:
\beq
\phi_{\text{Berry}}=i\oint\rm dk_z\bra{\Psi_{\vec B}}\frac{\partial}{\partial k_z}\ket{\Psi_{\vec B}};
\eeq 
here $\Psi_{\vec B}$ denotes the slater determinant of the 2D electron wave functions, for a given $k_z$ in the magnetic field. Analogous to the case in 1D, the derivative is understood to be taken only on the periodic part of the Bloch wave function. 

Despite that the integrand can be written as a sum over single-particle wave functions, we immediately notice that it cannot be expressed as a function of the single-particle density matrix. This is due to the fact that the wave function depends on the vector potential which is gauge dependent. One easy way to realize the fact is to consider a change of phase in the wave functions. The integrand is not invariant (the integral as a whole, on the other hand, is invariant modulo 2$\pi$) whereas the density matrix remain unaltered under the transformation.

We can, however, express the integral as a whole in terms of the density matrix by the following trick. The accumulated Berry's phase can be expressed as an integral of the Berry's curvature in one extra dimension using Stokes theorem, with the region of integration bounds by the original $k_z$ integral. The Berry's curvature can now readily be expressed in terms of the density matrix extended into the extra dimension, which is chosen continuously but otherwise aribitrarily with the constraint such that on the boundary, we have the original density matrix. We therefore have
\bea
i\oint_{\partial S}\rm dk_z\bra{\Psi_{\vec B}}\frac{\partial}{\partial k_z}\ket{\Psi_{\vec B}}&=&i\int_S\rm d^2k\epsilon^{\alpha\beta}\partial_\alpha\bra{\tilde\Psi_{\vec B}}\partial_\beta\ket{\tilde\Psi_{\vec B}}\nonumber\\
&=&i\int_S\rm d^2k\epsilon^{\alpha\beta}\Tr\left(\tilde\rho\partial_\alpha\tilde\rho\partial_\beta\tilde\rho\right);\nonumber\\
\label{bf3d}
\eea

$\ket{\tilde\Psi_{\vec B}}$ is the 2D electron many-body wave function, which in addition to being a function of $k_z$, has been extended to some extra direction $k_w$. $\tilde\rho=\sum_i\ket{\tilde\psi_{i}}\bra{\tilde\psi_{i}}$ is the extended 2D single particle density matrix, where $\ket{\tilde\psi_i}$ is a 2D single particle eigenstate in magnetic field $\vec B$. $\alpha, \beta$ run through two directions which is spanned by $k_z$ and $k_w$. The trace sums over both the band and the $(x,y)$-position basis. As mentioned above, the integral is chosen to be performed on the area such that the boundary is at $(k_z,k_w=0)$ and the density matrix has known values. It is straight forward to show the second equality, and the derivation is provided in the appendix. Different choices of density matrices inside the boundary can only alter the integral by multiples of $2\pi i$. To avoid cluttering of the equations, in the following we omit the tilde for the extended objects when there is no ambiguity.

Then following the formalism in Ref.\cite{AAJD}, we take the large size limit and expand $\rho$ to linear order in $B$. As discussed there, the density matrix in real space basis can be decomposed into two parts, one of which is translationally invariant:
\beq
\rho_{r_1,r_2}=\bar{\rho}_{r_1,r_2}\exp(-i\vec B\cdot(\vec r_1\times \vec r_2)/2),
\eeq
where $\rho_{r1,r2}$ denotes the density matrix in position basis, and $\bar\rho$ is translationally invariant. While the other part seems to be affected by the infinite range of $\bf r$, in our expression three $\rho$'s appear together and the combination is short-ranged and can be expanded in $B$. It is thus straight forward to expand $\rho$ explicitly and calculate. (Please see appendix for  details of the calculation.)

Up to first order in $B$, the result is
\bea
\phi_{\text{Berry}}&=&\int_S\mathrm d^2k\int_{BZ} \frac{\mathrm d^2k'}{(2\pi)^2}L_xL_y\epsilon^{\alpha\beta}\nonumber\\
&&\Big[\epsilon^{\gamma\delta}B\Tr(-\frac14\mathcal{F}_{\alpha\beta}
\mathcal{F}_{\gamma\delta}+\frac12\mathcal{F}_{\gamma\beta}
\mathcal{F}_{\alpha \delta})\nonumber\\
&&-i\partial_\alpha\Tr(\partial_\beta\rho_{0k'}(1-\rho_{0k'})\bar\rho_{k'}\rho_{0k'}-h.c.)\Big];
\label{keyresult}
\eea
$\alpha, \beta$ span $k_z, k_w$ and $\gamma, \delta$ span $k_x, k_y$. The integral of $k'$ is performed on the 2D Brillouin zone in $xy$-plane. $\bar\rho_{k'}=\bra{k'}\bar\rho\ket{k'}$ is the translationally invariant part of the density matrix at a given $(\vec k,k_w)$ and $\rho_{0k}$ is the density matrix in zero field. $\mathcal{F}_{\mu\nu}$ is the nonabelian Berry curvature of the occupied bands,
\beq
\mathcal{F}_{\mu\nu}=\partial_\mu\mathcal{A}_\nu-\partial_\nu\mathcal{A}_\mu-
i[\mathcal{A}_\mu,\mathcal{A}_\nu].
\eeq

Notice that in eqn. (\ref{keyresult}), the tensor structure in the first and the second term is different and we can rewrite the first term using the total-antisymmetric tensor in 4 dimensions:
\bea
\phi_{\text{Berry}}&=&\phi_I+\phi_A\nonumber\\
\phi_I&=&\frac{-\Phi_B}{32\pi^2}\int_{S\times BZ}\mathrm d^4{\bf k}\epsilon^{abcd}\Tr(\mathcal{F}_{ab}
\mathcal{F}_{cd})\\
\phi_A&=&\frac{-i\Phi_B}{4\pi^2}\int_{S\times BZ}\mathrm d^4{\bf k}\epsilon^{\alpha\beta}\partial_\alpha M_{\beta z}\\
M_{\alpha\beta}&=&\Tr\left(\partial_\alpha\rho_{0k'}(1-\rho_{0k'})\frac{\partial\rho_{k'}}{\partial B^\beta}\rho_{0k'}-h.c.\right).
\eea
We have explictly expanded the second term to first order in $\vec B$. $a, b, c, d$ runs through all directions. Both integrals are total derivatives and we can integrate back to the boundary which is the original 3D Brillouin zone:
\bea
\phi_I&=&\frac{-\Phi_B}{8\pi^2}\int\mathrm{ d}^3k\epsilon^{abc}\Tr(\mathcal{A}_a\partial_b\mathcal{A}_c-i\frac23\mathcal{A}_a
\mathcal{A}_b\mathcal{A}_c)\\
\phi_A&=&\frac{-i\Phi_B}{4\pi^2}\int\mathrm{d}^3k M_{zz}.
\eea
$\phi_I$ is isotropic, i.e., independent of the direction of the applied magnetic field. $\phi_A$, on the other hand, is anisotropic in the sense that if we do the same calculation for the magnetic field in $x$ or $y$ direction, the result in general would be different. Now we consider the gradual gauge transform in the $i$-direction and the magnetic field in the $j$-direction, the same calculation still goes through, provided that we take $\alpha, \beta$ in the $i$-direction and the extra direction, and $\gamma, \delta$ in the directions prependicular to the magnetic field. We get
\bea
\phi_{I,ij}&=&\phi_I\delta_{ij};\nonumber\\
\phi_{A,ij}&=&\frac{-i\Phi_B}{4\pi^2}\int\mathrm{d}^3kM_{ij}.
\eea
In terms of the effective theory, this means that the effective Lagrangian not only contains $\vec E\cdot\vec B$, in general we have $\sum_{ij}\alpha_{ij} E_iB_j$, where 
\beq
\alpha_{ij}=\int\frac{\mathrm{d}^3k}{(2\pi)^3}\left(\epsilon^{abc}\frac{-1}{2}\Tr(\mathcal{A}_a\partial_b\mathcal{A}_c-i\frac23\mathcal{A}_a
\mathcal{A}_b\mathcal{A}_c)\delta_{ij}+M_{ij}\right).
\eeq
By calculating the Berry's phase of these processes, not only do we get the coefficient of the topological term but we also get a part which is a physical response which agrees with Ref. \cite{AAJD}. In general $\sum_iM_{ii}$ also contributes to $\theta$. If TR symmetry is present then $M_{ij}=0$ and we see that the vacuum angle is shifted by $\pi$ in the presence of the strong topological insulator. We stress again that the calculation present here shows directly that the vacuum angle $\theta$ is shifted in the presence of the electronic band structure whereas the previous calculations only show that one can get current responses when one smoothly varies the Hamiltonian. Physically, our result predicts that there will be a half-charge electric flux if we put a strong topological insulator on a 3-torus with an odd number of magnetic flux, as described in the previous section; whereas in previous derivations, it is unclear if one can observe anything without either a boundary or a change of the Hamiltonian.

\section{Summary}
In this paper, we first show that there is a measurable difference for an effective theory containing either $\mathcal{L}_\theta$ or $\mathcal{L}_{\nabla\theta}$, in one and three spatial dimensions without a boundary. Specifically, with $\mathcal{L}_\theta$, in 1D there will be an electric field $\theta e/2\pi$ in the ground state. When $\theta=\pi$ the electric field can be in either direction and the CC symmetry and parity symmetry is spontaneously broken. In 3D, the same effect can be found on a 3-torus, but since it is the electric flux which is proportional to $\theta e$ in 3D, this effect vanishes in the thermaldynamic limit.

We then go on to show that the topological insulators in 1D and 3D can be described by $\mathcal{L}_\theta$ instead of $\mathcal{L}_{\nabla\theta}$. While the expression of $\theta$ agrees with previous results, to our knowledge this is the first derivation which distinguishes between $\mathcal{L}_\theta$ and $\mathcal{L}_{\nabla\theta}$.

We thank N. Nagaosa, X. G. Wen and F. Wilczek for helpful discussions. We acknowledge the support of NSF under grant DMR 0804040.

%\begin{acknowledgments}

% put your acknowledgments here.
%\end{acknowledgments}

%\end{enumerate}
\section{Appendix}
\textbf{A. Path integral formulation for the $\theta$-term}

In the main text, we derive the physical consequence of the $\theta$-term using the notion of $\theta$-vacuum, which is similar to a Hamiltonian formalism. One may wonder why we do not directly carry out the path integral. The first reason is that the quantization is not obvious if we just do the euclidean path integral as done below. The second reason is that if we calculate the fluctuation of the electric field at finite temperature, a naive calculation would give us a sum of {\it negative} values, which does not make sense. It turns out that for a free theory the position space path integral is ill-behaved and a positive finite term is expanded as an infinite negative sum. A similar situation occurs when one calculates the ground state energy of the bosonic string using mode expansion. Let we start right from the Lagrangian
\beq
\mathcal{L}=-\frac14 F^{\mu\nu}F_{\mu\nu}+\frac{e\theta}{2\pi}\epsilon^{\mu\nu}\partial_\mu A_\nu
\eeq
using the gauge $A_0=0$, for the $q=0$ sector at finite temperature $1/\beta$ we have the partition function
\beq
Z'_{q=0}=\left(\prod_{\omega_i}\left(\frac{2\pi}{\beta L\omega_i^2}\right)\right)\sum_ne^{in\theta}\exp\left(-\frac1{2\beta L}(\frac{2\pi n}{e})^2\right),
\eeq
where $\omega_i=2\pi i/\beta$ are the Matsubara frequencies. Again the finite frequency part decouples and the zero frequency part agrees with eqn.(\ref{z1d}) if we integrate $\ell$ first instead:
\beq
Z'_{q=0}\propto Z_{q=0}=\sum_ne^{in\theta}e^{-\frac1{2\beta L}(\frac{2\pi n}{e})^2}\equiv\sum_nW_n.
\eeq
Nevertheless, it is hard to see from this form that $\theta$ corresponds to a quantization condition for the electric field. Without reversing the $\ell$ integral, another way to see the $\theta$-dependence is to calculate the expectation value and the fluctuation of the electric field. For $\theta=\pi$, the expectation value would vanish and we can only rely on the fluctuation.

 If we calculate $\langle|\tilde{E^1}(0)|^2\rangle$, however, we would encounter a problem in the path integral as now all finite frequency part contributes and their sum seems to be infinitely negative:
\beq
\langle|\tilde{E^1}(0)|^2\rangle=\frac{L^2}{Z_0}\left(\sum_n-\left(\frac{2\pi n}{\beta Le}\right)^2W_n\right)-\sum_i\frac{2L}{\beta}.
\eeq
If we compare this to what we would have got using the Hamiltonian formalism,
\beq
\langle|\tilde{E^1}(0)|^2\rangle=\frac{L^2}{Z_0}\left(\sum_nW_n\left(-\left(\frac{2\pi n}{\beta Le}\right)^2+\frac{1}{\beta L}\right)\right).
\eeq
It seems we have to require 
\beq
\sum_i(1)=-\frac12.
\eeq
for the two expressions to agree. We can understand this equality by thinking of the left hand side as the zeta function at zero, $\zeta(0)$, written in a series. While the series is divergent at zero, the zeta function is well-defined and is indeed $-\frac12.$ 

The function $\sum_n n^2W_n$ is related to the elliptic $\Theta$-function. If one subtracts the fluctuation at $\theta=0$ from the expression and calculate at $\beta\rightarrow\infty$, one recovers that
\beq
\langle|\tilde{E^1}(0)|^2\rangle-\langle|\tilde{E^1}(0)|^2\rangle|_{\theta=0}
=L^2e^2\left(\frac{\theta}{2\pi}\right)^2,
\eeq
which implies the quantization.
\\

{\bf B. Derivation of eq. (\ref{bf3d}) and eq. (\ref{keyresult})}

Here we show explicitly the derivation of the second equality in eq. (\ref{bf3d}). First we note $\ket{\Psi_B}=\det(\psi_i)$, and 
\beq
\partial_\alpha\bra{\Psi_B}\partial_\beta\ket{\Psi_B}=
\sum_i\partial_\alpha\bra{\psi_i}\partial_\beta\ket{\psi_i}.\label{one}
\eeq
Now we plug in $\rho=\sum_i\ket{\psi_i}\bra{\psi_i}$ to the right hand side of the second equality, we have
\bea
&\Tr&(\rho\partial_\alpha\rho\partial_\beta\rho)\nonumber\\
&=&\sum_{ijk}\ket{\psi_i}\bra{\psi_i}\Big((\partial_\alpha\ket{\psi_i})\bra{\psi_i}+\ket{\psi_i}(\partial_\alpha\bra{\psi_i})
\Big)\nonumber\\
&&\Big((\partial_\beta\ket{\psi_i})\bra{\psi_i}+\ket{\psi_i}(\partial_\beta\bra{\psi_i})\Big)\nonumber\\
&=&\sum_{ij}\bra{\psi_i}\partial_\alpha\ket{\psi_j}\bra{\psi_j}\partial_\beta\ket{\psi_i}\nonumber\\
&&+\sum_i(\partial_\alpha\bra{\psi_i})(\partial_\beta\ket{\psi_i});\label{two}
\eea
in the derivation we have taken advantage of the fact that $\braket{\psi_i}{\psi_j} =\delta_{ij}$ and thus $(\partial_\alpha\bra{\psi_i})\ket{\psi_j}=-\bra{\psi_i}\partial_\alpha\ket{\psi_j}$.

Contract both eq. (\ref{one}) and eq. (\ref{two}) with $\epsilon^{\alpha\beta}$, we can see that they agree.

In the following we apply the magnetic field in the $z$-direction and take $\rho$ as a function of $k_z$, and $\vec r$ lies in the $xy$-plane. We take $\hbar=e=1$. $\bar\rho$ has the following matrix elements up to first order in $B$:
\bea
\bra{\psi_{nk}}\bar\rho\ket{\psi_{n'k'}}&=&\delta_{kk'}\left(\delta_{nn'}-
\frac14B\epsilon^{\gamma\delta}\mathcal{F}_{\gamma\delta,nn'}\right)\nonumber\\
\bra{\psi_{mk}}\bar\rho\ket{\psi_{m'k'}}&=&\frac14\delta_{kk'}
B\epsilon^{\gamma\delta}\check\mathcal{F}_{\gamma\delta,mm'}\nonumber\\
\bra{\psi_{nk}}\bar\rho\ket{\psi_{mk'}}&=&\delta_{kk'}\Big(\frac{i}{2}B\epsilon^{\gamma\delta}
\frac{\bra{\psi_{nk}}\{\partial_\gamma \rho_{0k},\partial_\delta H_k\}\ket{\psi_{mk}}}{E_{nk}-E_{mk}}\nonumber\\
&&+\frac{\bra{\psi_{nk}}H'_k\ket{\psi_{mk}}}{E_{nk}-E_{mk}}\Big);\label{me}
\eea
note that the momentum here is two-dimensional and everything has implicit $k_z, k_w$ dependence. $n$, $n'$ are indices for occupied bands and $m$, $m'$ are for empty bands. $\check\mathcal{F}$ is the nonabelian field strength for the Berry's phase gauge field defined from the unoccupied bands:
\bea
\check\mathcal{A}_{\mu,mm'}&=&-i\bra{u_{mk}}\frac{\partial}{\partial k^\mu}\ket{u_{m'k}}\nonumber\\
\check\mathcal{F}_{\mu\nu}&=&\partial_\mu\check\mathcal{A}_\nu-\partial_\nu\check\mathcal{A}_\mu
-i[\check\mathcal{A}_\mu,\check\mathcal{A}_\nu].
\eea
In the following computation one would find these expressions useful:
\bea
\mathcal{F}_{\mu\nu,nn'}&=&-i\sum_{m}\bra{\psi_{nk}}\partial_\mu\ket{\psi_{mk}}\bra{\psi_{mk}}
\partial_\nu\ket{\psi_{n'k}}\nonumber\\
&&-(\mu\leftrightarrow\nu);\nonumber\\
\check\mathcal{F}_{\mu\nu,mm'}&=&-i\sum_{n}\bra{\psi_{mk}}\partial_\mu\ket{\psi_{nk}}\bra{\psi_{nk}}
\partial_\nu\ket{\psi_{m'k}}\nonumber\\
&&-(\mu\leftrightarrow\nu);\nonumber\\
\epsilon^{\mu\nu\lambda\omega}\Tr(\mathcal{F}_{\mu\nu}\mathcal{F}_{\lambda\omega})&=&
\epsilon^{\mu\nu\lambda\omega}\Tr(\check\mathcal{F}_{\omega\mu}\check\mathcal{F}_{\nu\lambda}).
\label{gadget}
\eea
Note that in the expression for the Berry's curvature $\mathcal{F}$, we use the whole Bloch wave function $\ket{\psi}$ instead of the periodic part $\ket{u}$ but here it makes no difference.

Now we start from eqn.(\ref{bf3d}). Write explicitly in position basis, we have
\bea
\Tr\Big((\partial_\alpha\rho)&\rho&(\partial_\beta\rho)\Big)=\int\mathrm{d}r_1\mathrm{d}r_2\mathrm{d}r_3
(\partial_\alpha\bar\rho_{12})\bar\rho_{23}(\bar\partial_\beta\bar\rho_{31})\nonumber\\
&&\exp\left(-\frac{i}{2}B\epsilon^{\gamma\delta}(r_2-r_1)_\gamma(r_3-r_1)_\delta\right)\nonumber\\
&=&L_xL_y\int\frac{\mathrm{d}^2k'}{(2\pi)^2}\Big(\Tr\left((\partial_\alpha\bar\rho)\bar\rho(\partial_\beta\bar\rho)\right))\nonumber\\
&+&\frac{i}{2}B\epsilon^{\gamma\delta}\Tr\left((\partial_\alpha\partial_\gamma\rho_0)\rho_0(\partial_\delta
\partial_\beta\rho_0)\right)\Big)+\mathcal{O}(B^2),\nonumber\\\label{startp}
\eea
where in the second equality we taylor-expand in $B$, keep up to first order and go back to momentum space. We have also taken the infinite-size limit and make the sum of discrete momenta an integral. The trace on the right hand side traces over only the band indices.

The remaining task would be to plug in $\bar\rho$ and calculate explicitly to first order in $B$. One thing to notice is that when taking derivatives of $\bar\rho$, it acts not only on the matrix element but also on the basis. It is also useful to note that $\partial_u\rho_0$ only has non-vanishing matrix elements between the original occupied and empty states. 

As we can see from eqn.(\ref{me}), the inter-gap and intra-gap matrix element of $\bar\rho$ look pretty different. Let us denote the former as $\rho'$. $\rho'$ contributes only through the first term in the right hand side of eqn.(\ref{startp}); since $\rho'$ is already first order in $B$ the remaining $\bar\rho$ can be replaced by $\rho_0$. Let $\rho'=A+A^\dag$ with $A=(1-\rho_0)\rho'\rho_0$ (that is, $A$ is the matrix element connecting occupied bands to empty bands and vice versa for $A^\dag$), after explicit calculation, similar to eq. (\ref{two}), we have
\bea
&&\epsilon^{\alpha\beta}\left(\Tr(\partial_\alpha\rho'\rho_0\partial_\beta\rho_0)+
\Tr(\partial_\alpha\rho_0\rho'\partial_\beta\rho_0)+
\Tr(\partial_\alpha\rho_0\rho_0\partial_\beta\rho')\right)\nonumber\\
&=&-\partial_\alpha\left(\bra{n}\partial_\beta\ket{m}A_{mn}-c.c.\right)\nonumber\\
&=&\partial_\alpha\Tr\left(\partial_\beta\rho_0(1-\rho_0)\rho'\rho_0+h.c.\right);
\label{aniso}
\eea
$\ket{n}$ is the short hand notation of $\ket{\psi_{nk}}$ and repeated indices are summed over.

Now that the inter-gap matrix elements are dealt with, the remaining part of the first term can also be expanded and calculated:
\bea
\epsilon^{\alpha\beta}\Tr(\partial_\alpha\bar\rho\bar\rho\partial_\beta\bar\rho)|_{\text{remaining}}&=&
\epsilon^{\alpha\beta}\Big(\frac{i}2\Tr(\mathcal{F}_{\alpha\beta})\nonumber\\
&-&\frac{3i}{8}\epsilon^{\gamma\delta}\Tr\left(\mathcal{F}_{\alpha\beta}\mathcal{F}_{\gamma\delta}
-\check\mathcal{F}_{\alpha\beta}\check\mathcal{F}_{\gamma\delta}\right)\nonumber\\
&+&\mathcal{O}(B^2)\big).\label{part1}
\eea
The first term on the right hand side is proportional to $B^0$ and is similar to the polarization in 1D.

The only remaining part is the second term in eqn.(\ref{startp}). This part proves to be somewhat tricky to calculate as one has to manually group terms into expressions of $\mathcal{F}$ and $\check\mathcal{F}$. Nevertheless, it is otherwise straight forward and one gets
\bea
\epsilon^{\alpha\beta}&\epsilon^{\gamma\delta}&\Tr(\partial_\gamma\partial_\alpha\rho_0\rho_0\partial_\delta
\partial_\beta\rho_0)=\nonumber\\
&&\Tr\left(
\frac34\mathcal{F}_{\alpha\beta}\mathcal{F}_{\gamma\delta}
-\frac14\check\mathcal{F}_{\alpha\beta}\check\mathcal{F}_{\gamma\delta}
+\check\mathcal{F}_{\delta\alpha}\check\mathcal{F}_{\gamma\beta}
\right).\label{part2}
\eea
Combining eqn.(\ref{aniso}), eqn.(\ref{part1}), and eqn.(\ref{part2}), and with the help of eqn.(\ref{gadget}) we get egn. (\ref{keyresult}).

% Create the reference section using BibTeX:
%\bibliography{basename of .bib file}

%\end{fmffile}
\end{document}